\documentclass[twocolumn,aps,prb,reprint,noshowpacs,letterpaper,floatfix]%
{revtex4-1}
\usepackage{amsmath,amssymb}
\usepackage{txfonts}
\usepackage{bm}
\usepackage{textcomp}
\usepackage[pdftex]{graphicx}
\usepackage[bookmarks=false]{hyperref}
\begin{document}

\title{Spin-triplet superconductivity in the paramagnetic UCoGe under pressure
studied by $^{59}$Co NMR}

\author{Masahiro Manago}
  \email{manago@crystal.kobe-u.ac.jp}
  \altaffiliation[Present address: ]{Department of Physics, Kobe University,
  Kobe 657-8501, Japan}
\author{Shunsaku Kitagawa}
\author{Kenji Ishida}
  \affiliation{Department of Physics, Graduate School of Science,
  Kyoto University, Kyoto 606-8502, Japan}

\author{Kazuhiko Deguchi}
\author{Noriaki K. Sato}
  \affiliation{Department of Physics, Graduate School of Science,
  Nagoya University, Nagoya 464-8602, Japan}

\author{Tomoo Yamamura}
  \altaffiliation[Present address: ]{Institute for Integrated Radiation
  and Nuclear Science,
  Kyoto University, Kumatori 590-0494, Japan}
  \affiliation{Institute for Materials Research,
  Tohoku University, Sendai 980-8577, Japan}

\begin{abstract}
A $^{59}$Co nuclear magnetic resonance (NMR) measurement was performed on the
single-crystalline ferromagnetic (FM) superconductor UCoGe under a pressure of
1.09 GPa, where the FM state is suppressed and superconductivity occurs in the
paramagnetic (PM) state, to study the superconducting (SC) state in the PM
state.
$^{59}$Co-NMR spectra became broader but hardly shifted across the SC
transition temperature.
The Knight-shift change determined from fitting the spectral peak with a Gaussian
was much smaller than the spin part of the Knight shift; this is in good
agreement with the spin-triplet pairing suggested from the large upper critical
field.
The spectrum broadening in the SC state cannot be attributed to the SC
diamagnetic effect, but is related to the properties of spin-triplet
superconductivity.
The origins of the broadening are discussed herein.
\end{abstract}

\maketitle

\section{Introduction}
Unconventional superconductivity mediated by a mechanism other than ordinary
electron-phonon coupling is one of the most interesting phenomena in
condensed-matter physics, and the symmetry of the pairing often
differs from the $s$ wave.
Among the various unconventional superconductors reported so far, spin-triplet
(odd-parity) superconductors are quite rare systems.
Cooper pairs have spin
degrees of freedom in the superconducting (SC) state in this case, leading to
various exotic phenomena.
The spin-triplet pairing was first identified in superfluid
$^3$He \cite{RevModPhys.47.331}, and odd-parity pairing is most likely realized
in a heavy-fermion superconductor UPt$_3$ \cite{PhysRevLett.77.1374}.
The possibility of spin-triplet pairing was also noted in Sr$_2$RuO$_4$
\cite{Nature.396.658,RevModPhys.75.657,JPSJ.81.011009,NPJ.2.40}, and this system
has been intensively studied thus far to identify its pairing symmetry.

The family of ferromagnetic (FM) superconductors is another class of
unconventional superconductors \cite{JPSJ.81.011003,%
JPSJ.83.061011,JPSJ.76.051011,JPSJ.83.061012,PhysicaC.514.368}.
The odd-parity pairing is expected in this family because of the
coexistence of ferromagnetism and superconductivity.
The pairing mechanism of these systems is closely related to the FM
instability, and this pairing glue can be tuned by the external magnetic field,
resulting in the unusual field dependence of the SC transition temperature
$T_\text{SC}$.
For UCoGe, it was reported from a field-angle-controlled
nuclear magnetic resonance (NMR) experiment that the Ising-type FM fluctuations
along the $c$ axis are essential for the pairing \cite{PhysRevLett.108.066403}.
The changes in pairing strength with the fields were quantitatively
analyzed by measuring the upper critical field $H_\text{c2}$,
and this result supports the scenario of FM fluctuation-mediated
superconductivity \cite{NatComm.8.14480}.
FM fluctuations also play an important role in superconductivity in URhGe;
however, transverse fluctuations as well as those parallel to the $c$ axis were
also found to contribute to the pairing \cite{PhysRevLett.114.216401}.
This is confirmed by the enhancement of the superconductivity
by a uniaxial stress \cite{PhysRevLett.120.037001}.
Therefore, these systems are suitable for studying the SC
mechanism in detail.

UCoGe is a unique system because its SC phase remains beyond the suppression of
the FM phase by pressure \cite{JPSJ.77.073703,PhysRevLett.103.097003,%
PhysRevB.94.125110}, and its superconductivity is expected to possess
a spin-triplet symmetry on both the FM and paramagnetic (PM) sides.
Actually, $H_\text{c2}$ is far above
the Pauli-limiting field estimated from $T_\text{SC}$ at
ambient pressure \cite{PhysRevLett.100.077002} and under pressure above $P_c$
\cite{PhysRevLett.103.097003,PhysRevB.94.125110}
when the field is perpendicular to the $c$ axis.
These results are consistent with the spin-triplet pairing.
The absence of the Knight-shift decrease in the FM SC state also supports the
spin-triplet scenario \cite{JPSJ.83.061012}.
However, it is not trivial to achieve such a large $H_\text{c2}$ in the $a$
and $b$ axes because the magnetic easy axis is the $c$ axis and the spins of the
Cooper pair cannot rotate freely in the presence of strong spin-orbit coupling.
The absence of the Pauli-PM effect in the FM SC state is explained
by the presence of the large exchange field \cite{PhysRevB.81.180504};
however, this mechanism does not work on the PM side.
The rotation of the $\bm{d}$ vector perpendicular to the field is therefore
predicted under pressure with fields \cite{PhysRevB.93.174512}, where the
$\bm{d}$ vector has been conventionally used to represent the spin state in the
spin-triplet pairing and is perpendicular to the spins of the Cooper pairs.
Such a rotation of the $\bm{d}$ vector leads to an unchanged spin
susceptibility in the SC state, resulting in the absence of the Pauli-PM effect.
Because only a few experimental results on SC symmetry have been reported so far
in the PM SC state under pressure, the properties of this state remain unclear.

To obtain microscopic information about the superconductivity of UCoGe on the
PM side, we performed $^{59}$Co NMR Knight-shift measurements under pressure.
The result is in good agreement with the spin-triplet pairing suggested from the
large $H_\text{c2}$ when the field is applied to the $ab$ plane.
In this case, the NMR spectrum anomalously broadened, which cannot be understood
by the SC diamagnetic effect but is related to the
properties of spin-triplet superconductivity.

\section{Experiment}

\begin{figure}
    \centering
    \includegraphics[width=80mm]{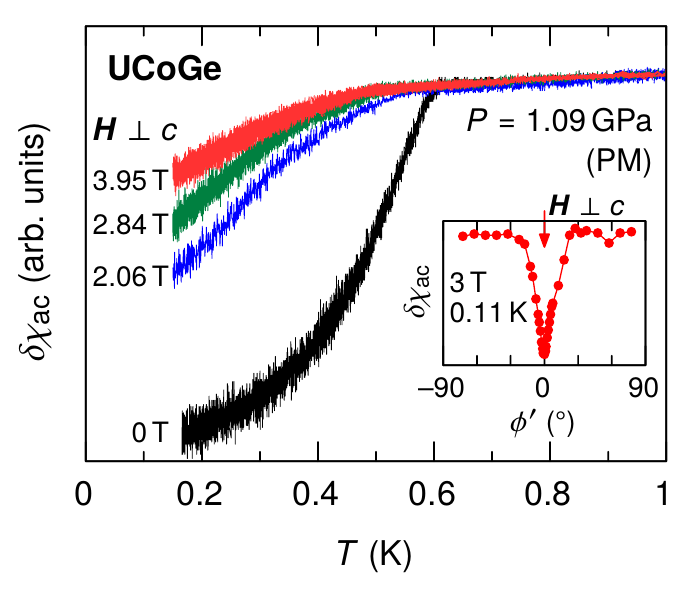}
    \caption{\label{fig:chiac}(Color online)
    Temperature dependence of ac susceptibilities $\delta \chi_{\text{ac}}$
    of UCoGe at some magnetic fields perpendicular to the $c$ axis
    as measured using an NMR coil on cooling.
    Superconducting transition is detected in all fields.
    The frequencies are 8--9 MHz.
    Inset: field-angle dependence of $\delta \chi_{\text{ac}}$ at 3 T
    and 0.11 K.
    }
\end{figure}

A single-crystalline sample was used for $^{59}$Co NMR measurements,
and its FM and SC transition temperatures at ambient pressure are
$T_\text{Curie} = 2.5$ K and $T_\text{SC} = 0.46$ K, respectively.
This sample is the same one used in a recent nuclear quadrupole resonance (NQR)
measurement under pressure \cite{PhysRevB.99.020506}.
Hydrostatic pressure was applied using a piston-cylinder-type cell with
Daphne oil 7373 as a pressure medium, and the applied pressure was
calibrated using an SC transition temperature of a Pb sample inside the cell.
The thermometer was attached outside the pressure cell.
Because the pressure cell and the thermometer are immersed in a
$^{3}$He-$^{4}$He mixture of a dilution refrigerator, thermal contact between
the sample and thermometer is sufficient for the present measurement.
The difference in $T_\text{SC}$ of UCoGe determined by the cooling and warming
processes is 0.01 K.
The magnetic field was applied using a transverse SC magnet,
enabling us to control the field angle precisely.

Figure \ref{fig:chiac} shows the temperature dependence of the ac
susceptibility $\delta \chi_{\text{ac}}$ of UCoGe at some magnetic fields with
the field perpendicular to the $c$ axis determined by the change of the tuning
frequency of the $LC$ circuit.
The SC transition temperature of UCoGe increases to $T_\text{SC} = 0.60$ K at
1.09 GPa.
No FM transition was detected at this pressure \cite{PhysRevB.99.020506}.
The field was aligned from the local minimum of the field-angle
dependence of $\delta \chi_{\text{ac}}$ at $T=0.11$ K with 3-T field,
as shown in the inset of Fig.~\ref{fig:chiac}.
A clear diamagnetic signal was observed in the narrow-angle region
with $H \perp c$ because of the large $H_\text{c2}$ anisotropy.
The field direction in the $ab$ plane could not be determined only from
$\delta \chi_{\text{ac}}$, and this was identified from the NMR spectra,
as shown below.

The crystal structure of UCoGe possesses a $\mathit{Pnma}$
space group (No.~62, $D_{2h}^{16}$), and the Co site only has mirror
symmetry with respect to the $b$ plane \cite{JAlloysCompd.234.225}.
The Co sites will split into two sites if the external magnetic field is not
parallel to the $\mathit{ab}$ or $\mathit{bc}$ plane for the NMR measurements.
The single site is observed with fields parallel to the $\mathit{ab}$ plane,
but the spectral positions differ in different
field directions owing to the low symmetry of the local site.
This feature enables us to deduce the full
information about the applied field from the NMR spectrum.

The nuclear Hamiltonian of $^{59}$Co ($I=7/2$) in UCoGe consists of two parts.
One is a Zeeman Hamiltonian of the nuclear magnetic moment,
and the other is the electric quadrupole Hamiltonian arising from the coupling
between the nuclear quadrupole moment and the electric field gradient (EFG).
Then, the total Hamiltonian is represented as
\begin{align}
\mathcal{H} &= \mathcal{H}_{Z} + \mathcal{H}_{Q} \notag\\
&= -\gamma_N \hbar (1 + \bm{K}) \bm{I} \cdot \bm{H} \notag\\
&\phantom{=}{}+\frac{\hbar \omega_Q}{6} \left\{
[3I_z^2 -I(I+1)] + \frac{1}{2} \eta (I_{+}^2 + I_{-}^2)
\right\},
\end{align}
where $\gamma_N$ is a gyromagnetic ratio of the nucleus,
$\bm{H}$ is an external magnetic field, $\bm{K}$ is a Knight-shift tensor,
$\omega_Q$ is a quadrupole frequency, and $\eta$ is an asymmetric
parameter of the EFG\@.
This Hamiltonian is an expression in a particular coordinate, namely,
the principal axis of the EFG\@.
The $z$ axis is the direction where the EFG is maximum, and the $y$ is the second
maximum direction.
In the case of the $^{59}$Co site in UCoGe,
the $z$ direction is $\sim 10$\textdegree\ tilted to the crystallographic
$c$ axis from the $a$ axis, and the $y$ axis is parallel to the $b$ axis
\cite{PhysRevLett.105.206403}.
The quadrupole parameters are $\nu_Q \equiv \omega_Q/(2\pi) = 2.85$ MHz and
$\eta=0.52$ at ambient pressure \cite{JPSJ.79.023707}, and these values
slightly change as pressure increases.
We used the values $\nu_Q = 2.795$ MHz and $\eta=0.535$
determined at 1.09 GPa \cite{PhysRevB.99.020506}.
If the Zeeman Hamiltonian is a dominant term in the above Hamiltonian,
seven lines will be observed, and they arise from $m \leftrightarrow m-1$
transitions ($m=7/2, 5/2, \dots, -5/2$).
These line positions strongly depend on the field angle with respect to the EFG
coordinate as well as the anisotropy of the Knight shift, and thus, the
analysis of the spectral positions enables us to deduce the field direction.
The Knight shift was measured at the central line arising from the
$1/2 \leftrightarrow -1/2$ transition, and $^{59}\gamma_N / 2\pi = 10.03$ MHz/T
was used as a nuclear gyromagnetic ratio.
The effect of the quadrupole shift was subtracted by using the numerical
diagonalization of the nuclear Hamiltonian.

\begin{figure}
    \centering
    \includegraphics[width=80mm]{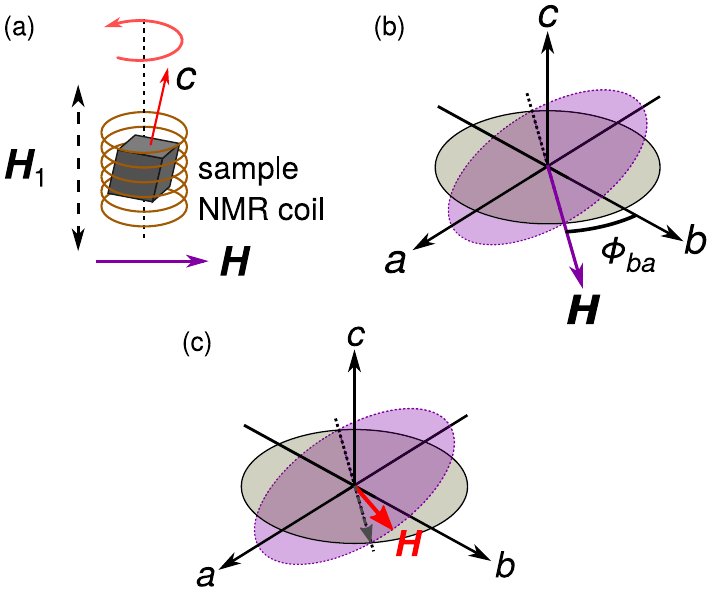}
    \caption{\label{fig:nmr-field}(Color online)
    (a) The schematic image of the static field $H$, NMR rf field $H_1$,
    and the sample directions.
    (b) The field-angle direction.
    The magnetic field is not perpendicular to the $c$ axis in most directions
    and is perpendicular to the $c$ axis only at a point.
    (c) The field slightly tilted toward the $c$ axis.
    }
\end{figure}

\begin{figure}
    \centering
    \includegraphics[width=80mm]{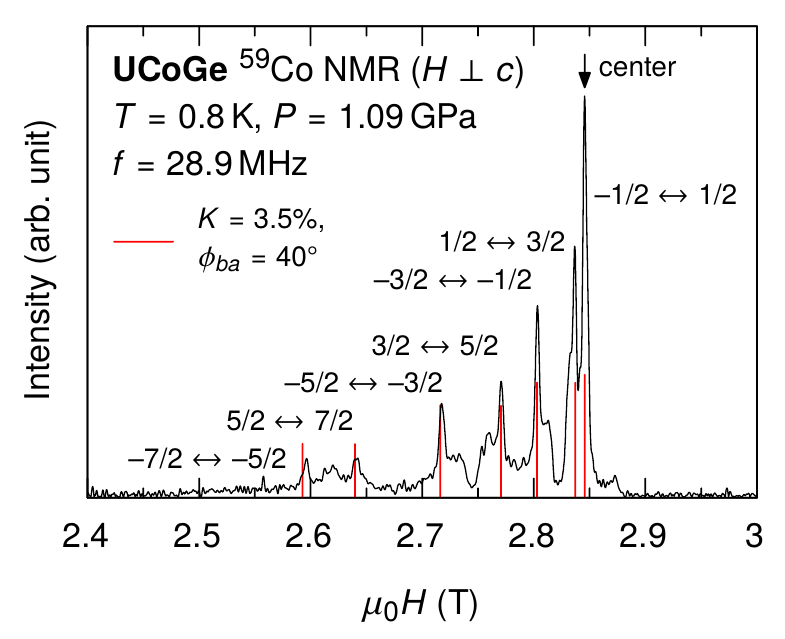}
    \caption{\label{fig:hsp}(Color online)
    Field-swept $^{59}$Co NMR spectra with the field perpendicular to the $c$
    axis at 0.8 K (normal state).
    The vertical red line is the best fit of the spectra with
    $\phi_{ba}=40$\textdegree\ by numerical diagonalization of the nuclear
    Hamiltonian.
    The vertical black arrow indicates the ``central''
    ($1/2 \leftrightarrow -1/2$) peak, at which the Knight shift was measured.
    This peak is not located at the center of the seven splitting lines owing
    to the low-symmetric field direction.
    }
\end{figure}

An attempt was made to attach the sample inside the pressure cell such that the
crystallographic $c$ axis is along the vertical direction;
however, it was found that the $c$ axis was tilted.
Because we could rotate the field angle only horizontally,
the field direction was not aligned parallel to the $a$ or $b$ axis.
The schematic image of the field alignment for the NMR measurements
is shown in Fig.~\ref{fig:nmr-field}.
This field direction perpendicular to the $c$ axis was identified from the
analyses of the field-swept NMR spectra at 0.8 K in the normal state, as
shown in Fig.~\ref{fig:hsp}.
Seven peaks were observed, which are due to the nuclear quadrupole splitting
of $^{59}$Co and consistent with $H \perp c$,
although non-negligible broadening was visible at the satellite peaks.
The vertical red line is the best fit of the calculated NMR peak positions
by numerical diagonalization.
The field angle tilts by $\phi_{ba} =40$\textdegree\ from the $b$ axis to the
$a$ axis.
Further measurement of the NMR spectrum with a different field direction
revealed that the $c$ axis of the sample is tilted by 17\textdegree\ from
the vertical direction.
See Appendix A for the details on determining this angle.

\section{Results and discussion}

\begin{figure*}
    \centering
    \includegraphics{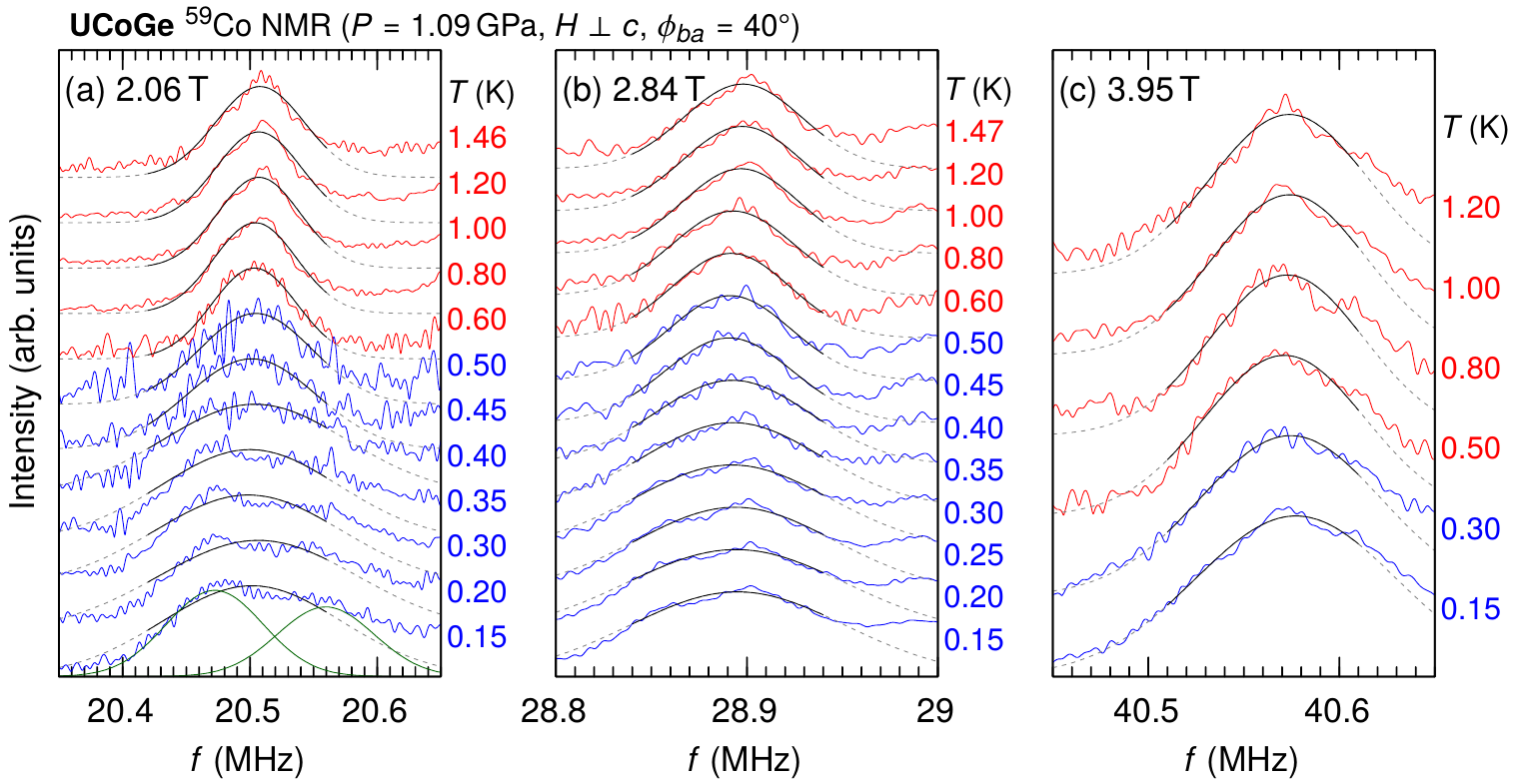}
    \caption{\label{fig:spectrum}(Color online)
    $^{59}$Co NMR spectra with field perpendicular to the $c$ axis.
    The solid black lines indicate the fitted result obtained using a
    Gaussian and represent the fitting range.
    The range was shifted lower to avoid the effect of the first satellite
    peak at higher frequency near the central peak.
    The red (blue) lines represent the result of the normal (SC) state.
    The spectra were also fitted with two-component Gaussians at 2.06 T in
    the SC state, and each part is shown for 0.15 K with green lines
    (see the Appendix for details).
    }
\end{figure*}

\begin{figure}
    \centering
    \includegraphics[width=80mm]{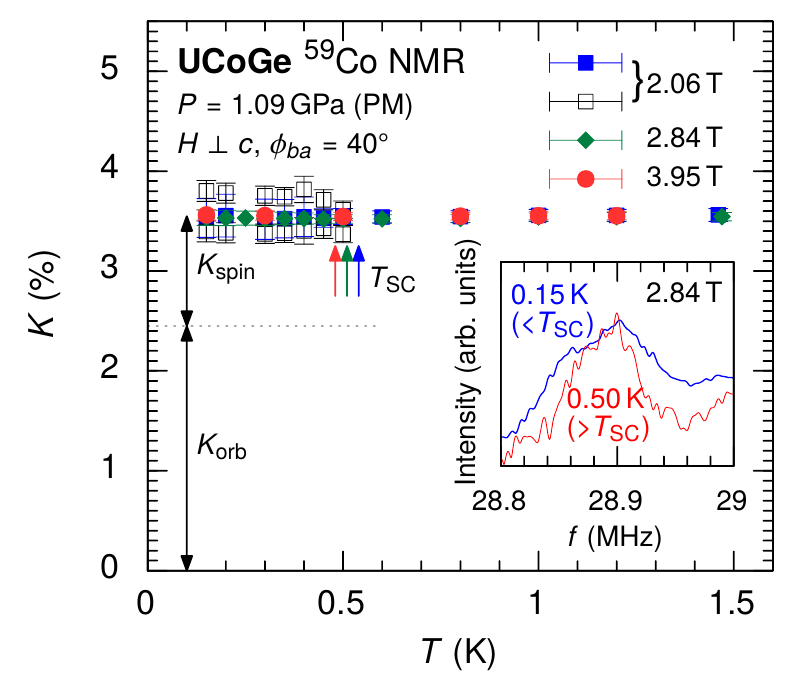}
    \caption{\label{fig:knightshift}(Color online)
    Temperature dependence of the $^{59}$Co NMR Knight shift of UCoGe
    with the field $40$\textdegree\ away from the $b$ axis to the $a$ axis
    at the central ($1/2 \leftrightarrow -1/2$) line.
    The errors are estimated from the 90\% (80\% for the SC state at 2.06 T)
    width of the spectra.
    The open black squares indicate the result of the two-component fitting
    in the SC state (see text for details).
    Inset: $^{59}$Co-NMR spectra at the central line at 2.84 T.
    Another peak at around 29 MHz originates from the first satellite owing to
    quadrupole splitting.
    }
\end{figure}

\begin{figure}
    \centering
    \includegraphics[width=80mm]{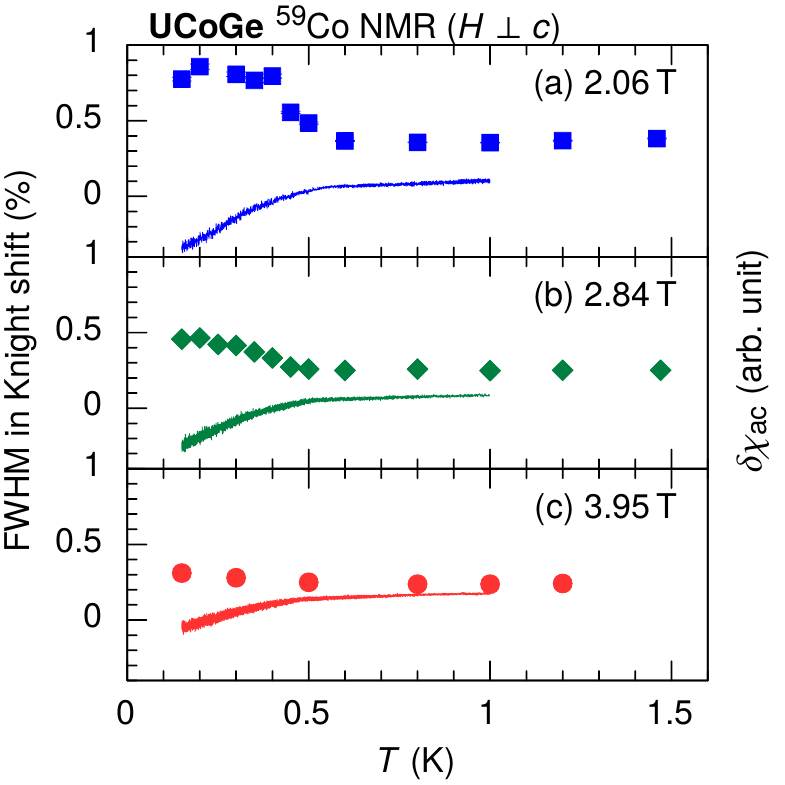}
    \caption{\label{fig:width}(Color online)
    Temperature dependence of the full width at half maximum (FWHM) of
    $^{59}$Co NMR spectra at the central line represented in terms of the
    Knight-shift distribution.
    The ac susceptibilities are shown for comparison.
    }
\end{figure}

Figure \ref{fig:spectrum} shows the $^{59}$Co NMR spectra with a field of 2--4 T
parallel to the $ab$ plane,
sufficiently smaller than $\mu_0 H_\text{c2} \sim$ 15 T
\cite{PhysRevLett.103.097003,PhysRevB.94.125110}.
The Knight shift was analyzed by fitting the spectral peak
with a Gaussian function, and the result is shown in Fig.~\ref{fig:knightshift}.
The Knight shift is expressed as $K(T) = K_\text{spin}(T) + K_\text{orb}$,
where $K_\text{spin}(T)$ is proportional to temperature-dependent spin
susceptibility and the orbital shift $K_\text{orb}$ is usually temperature
independent.
The spin and orbital components of the $^{59}$Co site in UCoGe are evaluated
from the $^{59}$Co and $^{73}$Ge NMR at ambient pressure
\cite{PhysRevB.97.075130}.
The spin component $K_\text{spin}$ along the $b$ axis
is not far from the
estimated value from the experimental specific-heat coefficient
in the normal state and hyperfine coupling constant \cite{JPSJ.74.1245}.
The spin component is $K_\text{spin} \simeq 1.1$\% with this field
direction at low temperatures below 1 K, as shown by the vertical arrow in
Fig.~\ref{fig:knightshift}, and it mainly originates from the spin
susceptibility along the $b$ axis because of the in-plane magnetic anisotropy
\cite{PhysRevB.97.075130}.
The observed change across $T_\text{SC}$ is an order of 0.1\% and is
much smaller than the spin part of the Knight shift.
This result indicates that the spin susceptibility $\chi_\text{spin}$ does not
decrease in the SC state drastically, and the possibility of spin-singlet
pairing is excluded, and spin-triplet superconductivity is suggested
in this phase.
We note that the NMR spectra became appreciably broader across $T_\text{SC}$,
as shown in the inset of Fig.~\ref{fig:knightshift} and in Fig.~\ref{fig:width};
this anomaly indicates that the Knight shift is actually measured 
in the SC state.
In ordinary superconductors, this broadening is attributed to the
effect of the vortex; however, the diamagnetic shift in
the vortex state of UCoGe is estimated to be $\sim 10^{-4}$\% at 2 T
if we use parameters at ambient pressure \cite{JPSJ.79.083708}, and
therefore, the broadening originates from the change in the spin
susceptibility.

\begin{figure}
    \centering
    \includegraphics[width=80mm]{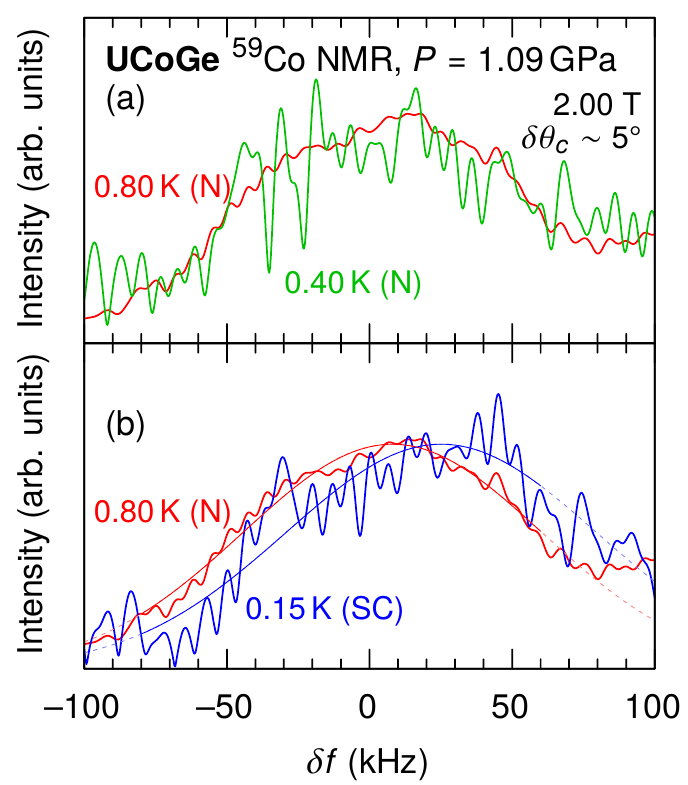}
    \caption{\label{fig:spectrum-tilt}(Color online)
    $^{59}$Co NMR spectra with 2.00-T field tilted to the $c$ axis by
    $\delta \theta_c \sim 5$\textdegree\ at 20.5 MHz.
    (a) The spectra in the normal state (0.80 and 0.40 K).
    No large shifts are observed in the normal state at these temperatures.
    (b) The spectra in the normal and SC states (0.80 and 0.15 K).
    The NMR spectrum slightly shifts to higher frequency in the SC state.
    The solid curves indicate results fitted with a Gaussian.
    }
\end{figure}

To reveal the origin of the anomalous broadening of the NMR spectra, the
spectra were measured with a tilted field.
Figure \ref{fig:spectrum-tilt} shows the NMR spectra with the field tilted by
$\delta \theta_c \sim 5$\textdegree\ from the
$ab$ plane to the $c$ axis.
The projected field to the $ab$ plane is $\phi_{ba} \simeq 24$\textdegree\ away
from the $b$ axis to the $a$ axis.
The Knight shift showed a small increase in the SC state without any significant
broadening, as shown in Fig.~\ref{fig:spectrum-tilt}.
This indicates that the broadening of the spectrum in the $H \parallel ab$
plane arises from the following two effects.
One is the Knight-shift decrease from the sample region where the field is
exactly aligned in plane, and the other is the Knight-shift increase arising
from the tiny misalignment of the field from the $c$ axis.
This misalignment is smaller than $\pm 5$\textdegree.

It is surprising that the NMR spectra in the SC state exhibit a drastic change
with tilting the field by only $\sim 5$\textdegree\ to the $c$ axis.
This result originates from the high sensitivity of UCoGe to the $c$ component of
the field.
Such an anomalous field-angle dependence is also seen at the ambient pressure
in $H_\text{c2}$ \cite{JPSJ.78.113709} and nuclear spin-lattice relaxation rate
$1/T_1$ \cite{PhysRevLett.108.066403}.
This feature is considered to originate from the strong Ising anisotropy
along the $c$ axis of UCoGe, and the application of the field along this
direction changes the electronic state.

The origin of this field-angle distribution is most likely due to the
sample distortion by the pressure.
The details of this discussion and the analysis based on the two-component
Gaussian with $H \parallel ab$ are shown in Appendix B.
The result of the two-component fitting at 2.06 T is shown in
Fig.~\ref{fig:knightshift}, and the increasing and decreasing parts are
reasonably extracted in the SC state.
Although clear two-component spectra are not seen at higher fields,
the broadening of the spectra in the SC state suggests that
two-component spectra persist even at higher fields.
The Knight-shift decreases in the SC state for a field of 2--4 T are of the
order of $10^{-1}$\%; this change corresponds to $\sim 10^{-1}\chi_\text{spin}$.
The increasing Knight shift is also of the order of $10^{-1}$\%.

It is not trivial that the Knight shift increases in the SC state in the
spin-triplet pairing.
A possible origin of the increase in spin susceptibility is the sharp density
of states (DOS) around the Fermi energy \cite{JPSJ.83.053701}.
If the DOS has a large slope at $E_\text{F}$, a larger energy gain is obtained in
the SC state, and the spin-triplet Cooper pairs have additional spin
polarization.
Although the details of the DOS in the normal state are not fully understood
in UCoGe, a peak structure around $E_\text{F}$ was revealed by a photoemission
spectrum \cite{PhysRevB.91.174503}, and this could lead to redistribution
of the Cooper pair spins.
This effect would be most prominent when the field is along the easy axis
($c$ axis), consistent with the experimental results.
Another possibility is that the metamagnetism occurs coincident with
the SC transition when the field has the $c$-axis component.
A metamagnetic behavior at low temperatures is predicted using the theory of the
quantum itinerant ferromagnetism \cite{RevModPhys.88.025006}.
This effect would also be anisotropic and sensitive to the $c$-axis field.
We note that an increase of the Knight shift below 1 K was also
observed in the FM SC state of UCoGe in the present sample at ambient pressure
\cite{JPSJ.83.061012}; however, this anomaly started to occur at a temperature
far above $T_\text{SC}$.
Such an increase in the Knight shift in the normal state was not observed in
the PM state under pressure.
Thus, we attribute this anomaly to the existence of the FM phase.

The decreasing Knight shift in the SC state is interpreted as the (partial)
pinning of the $\bm{d}$ vector in the case of spin-triplet pairing.
This result is in contrast to the case of the FM SC state,
in which no decrease of the Knight shift was detected within the experimental
resolution when $H \parallel a$ and $b$ \cite{JPSJ.83.061012}.
The almost constant spin susceptibility in the FM SC state
in these directions is attributed to the presence of the exchange field,
which weakens the Pauli-PM effect \cite{PhysRevB.81.180504}.
It should be noted that exchange field is absent in the PM SC state under
pressure.
This suggests that the $\bm{d}$ vector would be perpendicular to the $b$ axis
in the $ab$ plane and presumably parallel to the $a$ axis, at least in the
high-field region because the small spin susceptibility along the $a$ axis
\cite{PhysRevB.97.075130} is in the favor of $\bm{d} \parallel a$.
Thus, it is crucial to reveal whether the $\bm{d}$ vector rotates with the
field parallel to the $ab$ plane in the PM SC state.
The field dependence of the broadening of the NMR spectra in the SC state
implies that the decreasing component becomes weaker at higher field.
This result may originate from the rotation of the $\bm{d}$ vector by the field.
However, it is difficult to distinguish whether this field dependence is mainly
ascribed to the gradual rotation of the $\bm{d}$ vector or the quasiparticle
excitation from the present result.
Because the possibility of multigap superconductivity is noted from
thermal transport measurements in the FM SC state \cite{PhysRevB.90.180501},
it is inappropriate to speculate on the quasiparticle excitation based on a simple
SC gap model.
To evaluate the quasiparticle contribution accurately,
it is important to reveal the SC gap structure under the magnetic field
through, for instance, specific heat measurements,
and future experiments may reveal the details of the $\bm{d}$-vector rotation.

Another interesting problem concerning the Knight-shift decrease is the
decreasing amount of the Knight shift in the SC state, which
is somewhat smaller than the simple estimate in the case of perfect pinning.
If the $\bm{d}$-vector rotation occurs for a field much smaller than 2 T,
the Knight shift becomes (almost) constant below $T_\text{SC}$.
The existence of a residual DOS in the present sample may cause the reduction of
the decreasing $\chi_\text{spin}$ as observed in $1/T_1$ without a field
\cite{PhysRevB.99.020506}.
We also note that the value of $K_\text{orb}$ is assumed to be unchanged
by pressure when the spin part of the Knight shift is estimated.
If $K_\text{orb}$ changes under pressure, this could cause
overestimation of the spin part.
A more intrinsic reason for the small decrease of the Knight shift is also
anticipated in heavy-fermion spin-triplet superconductors:
it was reported that the Knight-shift decrease is much smaller than the
expected $\chi_\text{spin}$ in UPt$_3$ even if the $\bm{d}$ vector is believed
to be pinned along the field direction \cite{PhysRevLett.80.3129}.
At present, this phenomenon remains unexplained.

\section{Conclusion}

We demonstrated that the $^{59}$Co NMR spectra are hardly shifted but become
broader in the SC state of the single-crystalline UCoGe under a
pressure of 1.09 GPa in the PM state when the field is perpendicular to the
$c$ axis.
This result indicates that the spin-triplet pairing is realized in the PM state.
The small decreasing component suggests the pinning of the $\bm{d}$ vector
in the $ab$ plane.
Furthermore, the magnetic field dependence implies that the pinning is not so
strong that the $\bm{d}$ vector can rotate to avoid the Pauli-PM effect;
this is consistent with the large $H_\text{c2}$ perpendicular to the $c$ axis.
An increasing Knight shift was also detected, arising from the sample
region with a tilted field to the $c$ axis.
These anomalies cannot be interpreted by a spin-singlet pairing at all
but are indicative of spin-triplet pairing.

\begin{acknowledgments}
The authors would like to thank Y. Tokunaga, T. Hattori, D. Aoki, Y. Maeno,
S. Yonezawa, A. Daido, Y. Yanase, J.-P. Brison, D. Braithwaite, A. Pourret,
C. Berthier, A. de Visser, J. Flouquet, and V. P. Mineev for valuable
discussions.
This work was supported by the Kyoto University LTM Center,
Grants-in-Aid for Scientific Research
(Grants No.~JP15H05745 and No.~JP17K14339),
Grants-in-Aid for Scientific Research on Innovative Areas ``J-Physics''
(Grants No.~JP15H05882, No.~JP15H05884, and No.~JP15K21732), and a
Grant-in-Aid for JSPS Research Fellows (Grant No.~JP17J05509) from JSPS.
\end{acknowledgments}

\appendix

\section{Details of the sample tilting}

\begin{figure}
    \centering
    \includegraphics[width=80mm]{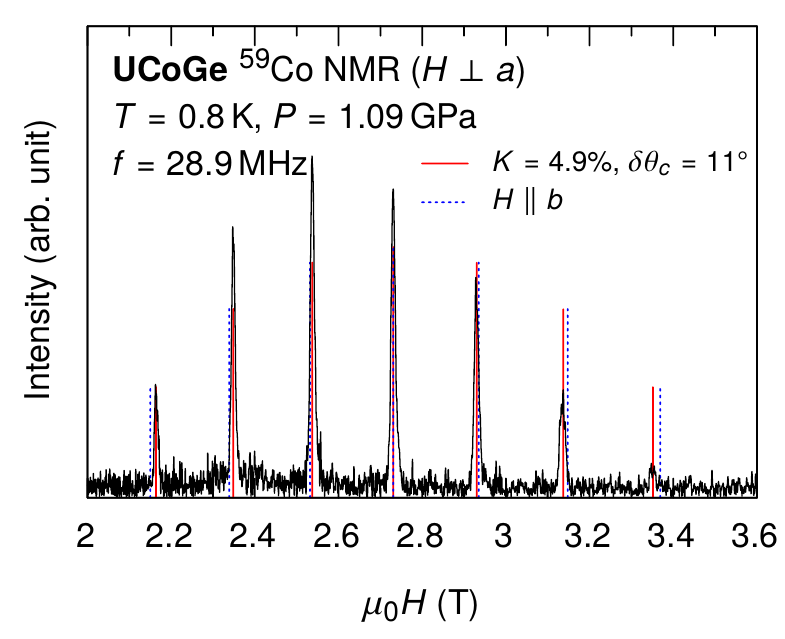}
    \caption{\label{fig:hsp-bc}(Color online)
    Field-swept $^{59}$Co NMR spectra with the field perpendicular to the $a$
    axis at 0.8 K (normal state).
    The vertical solid red lines indicate the best fit
    of the spectra with $\delta \theta_{c}=11$\textdegree.
    The dashed blue lines indicate the calculated result with $H \parallel b$.
    }
\end{figure}

We estimate how much the sample is tilted inside the cell.
Figure \ref{fig:hsp-bc} shows the NMR spectrum with $H \perp a$.
The field was aligned by monitoring the splitting of the central line: if the
field is not parallel to the $\mathit{bc}$ plane, the central peak splits into
two lines owing to the low-symmetry nature of the Co site.
This field direction is close to the $b$ axis but is tilted
toward the $c$ axis due to the sample misalignment.
The dashed blue lines indicate the calculated result with $H \parallel
b$ and do not fit with the experimental result.
The best fit was obtained when the field was tilted
by $\delta \theta_c \sim 11$\textdegree\ from the $b$ axis to the $c$ axis,
as shown by the solid red vertical lines.
Then, it was revealed that the $c$ axis of the sample is tilted by
17\textdegree\ from the vertical direction from this result combined with
the field angle with $H \perp c$.
This information was used to rotate the field by
$\delta \theta_c \sim 5$\textdegree\ toward the $c$ axis.
There are two possible directions with $\delta \theta_c \sim 5$\textdegree,
and the tilted field is closer to the $b$ axis than the original direction
with $H \perp c$, as shown in Fig.~\ref{fig:nmr-field}(c).

\section{Two-component Knight shifts and their origin}

\begin{figure}
    \centering
    \includegraphics[width=80mm]{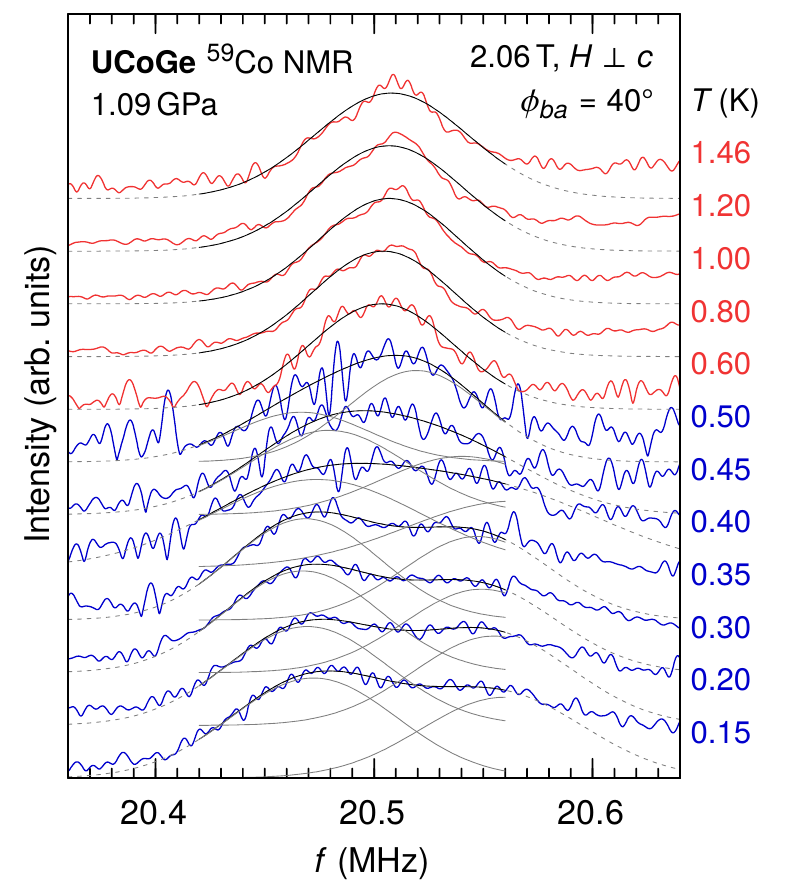}
    \caption{\label{fig:nmr-spectrum-2T-full}(Color online)
    $^{59}$Co NMR spectra with the field perpendicular to the $c$ axis,
    analyzed by two Gaussians in the SC state.
    The experimental data are the same as the ones shown in
    Fig.~\ref{fig:spectrum}(a).
    The solid black lines indicate the best fit with the Gaussian(s).
    The solid gray lines indicate the partial components of two-component
    fitting.
    }
\end{figure}

The NMR spectra at 2.06 T were also analyzed based on the two-component Gaussian.
The spectra and the fitting results are shown in
Fig.~\ref{fig:nmr-spectrum-2T-full}.
This analysis indicates that both the increasing and decreasing parts exist
in the SC state at 2.06 T.
They are on the order of $\Delta K \sim 0.1$\%, which is clear but much smaller
than the spin part of the Knight shift $K_\text{spin} \simeq 1.1$\%.
We could not extract two components separately at higher fields owing to the
smaller splitting.

It is not trivial that the $^{59}$Co NMR Knight shift has two components in
the SC state in the single-crystalline sample.
We consider that this anomaly is due to the sample distortion by the
pressure based on the following experimental results.

\begin{figure}\centering
    \includegraphics[width=80mm]{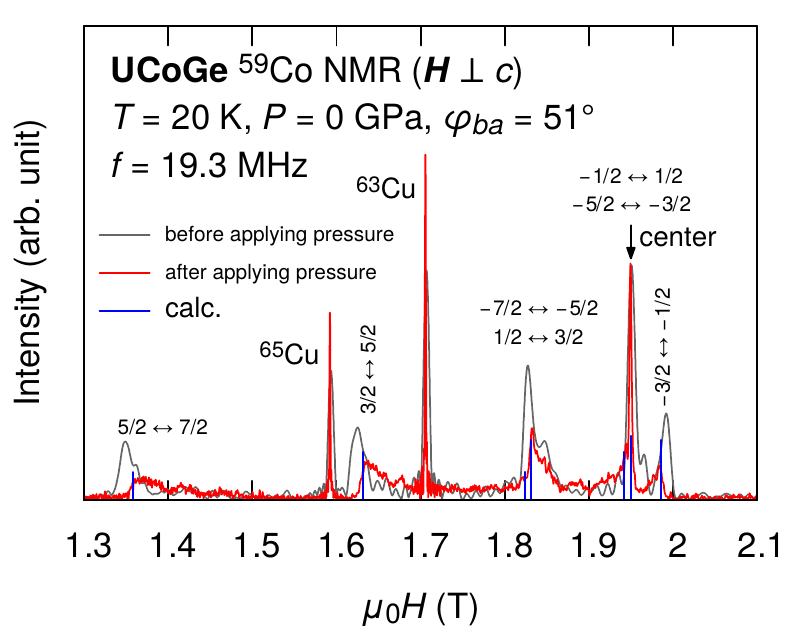}
    \caption{\label{fig:nmr-spectrum-0GPa}(Color online) Field-swept
    $^{59}$Co NMR spectra at 0 GPa at 20 K at the PM state
    before (gray) and after (red) applying pressure.
    The vertical blue line is the best fit
    of the spectra with $\phi_{ba}=51$\textdegree.
    }
\end{figure}

\begin{figure}\centering
    \includegraphics[width=80mm]{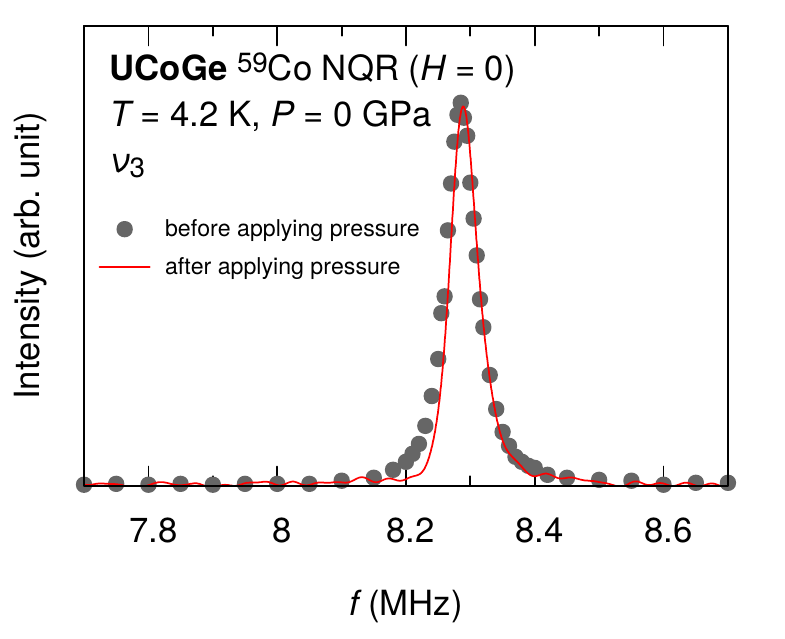}
    \caption{\label{fig:nqr-spectrum-0GPa}(Color online)
    $^{59}$Co NQR spectra without field at 0 GPa at 4.2 K (the PM state)
    before (gray circles) and after (red line)
    applying pressure.
    }
\end{figure}

Field-swept NMR spectra were measured at ambient pressure in the PM
state at 20 K before and after applying the pressure, as shown in
Fig.~\ref{fig:nmr-spectrum-0GPa}.
The field was applied perpendicular to the $c$ axis and was tilted
51\textdegree\ away from the $b$ axis to the $a$ axis.
This field direction differs from the one under the pressure measurement due to
inevitable tilting of the sample inside the pressure cell (see Fig.~\ref{fig:nmr-field}).
We found that the satellite peaks were broader after applying the pressure
than before applying the pressure, while the central peaks did not broaden.
Because the satellite peaks are more sensitive to the change in the EFG tensor,
this broadening implies inhomogeneous EFG with respect to the field direction;
more concretely, the quadrupole frequency $\nu_Q$ and/or the field angle with
respect to the crystal distributes even after removing the pressure.
However, the NQR spectrum at zero field at 4.2 K (PM state) did not
broaden within the experimental error, as shown in
Fig.~\ref{fig:nqr-spectrum-0GPa}.
This result means that $\nu_Q$ and its distribution did not change
after applying the pressure.
Therefore, it is likely that the sample is distorted and the field angle
distributes with respect to the crystal.
We consider that this distortion was induced by applying pressure and
existed at 1.09 GPa.

% \bibliography{bibliography}

\begin{thebibliography}{34}%
\makeatletter
\providecommand \@ifxundefined [1]{%
 \@ifx{#1\undefined}
}%
\providecommand \@ifnum [1]{%
 \ifnum #1\expandafter \@firstoftwo
 \else \expandafter \@secondoftwo
 \fi
}%
\providecommand \@ifx [1]{%
 \ifx #1\expandafter \@firstoftwo
 \else \expandafter \@secondoftwo
 \fi
}%
\providecommand \natexlab [1]{#1}%
\providecommand \enquote  [1]{``#1''}%
\providecommand \bibnamefont  [1]{#1}%
\providecommand \bibfnamefont [1]{#1}%
\providecommand \citenamefont [1]{#1}%
\providecommand \href@noop [0]{\@secondoftwo}%
\providecommand \href [0]{\begingroup \@sanitize@url \@href}%
\providecommand \@href[1]{\@@startlink{#1}\@@href}%
\providecommand \@@href[1]{\endgroup#1\@@endlink}%
\providecommand \@sanitize@url [0]{\catcode `\\12\catcode `\$12\catcode
  `\&12\catcode `\#12\catcode `\^12\catcode `\_12\catcode `\%12\relax}%
\providecommand \@@startlink[1]{}%
\providecommand \@@endlink[0]{}%
\providecommand \url  [0]{\begingroup\@sanitize@url \@url }%
\providecommand \@url [1]{\endgroup\@href {#1}{\urlprefix }}%
\providecommand \urlprefix  [0]{URL }%
\providecommand \Eprint [0]{\href }%
\providecommand \doibase [0]{http://dx.doi.org/}%
\providecommand \selectlanguage [0]{\@gobble}%
\providecommand \bibinfo  [0]{\@secondoftwo}%
\providecommand \bibfield  [0]{\@secondoftwo}%
\providecommand \translation [1]{[#1]}%
\providecommand \BibitemOpen [0]{}%
\providecommand \bibitemStop [0]{}%
\providecommand \bibitemNoStop [0]{.\EOS\space}%
\providecommand \EOS [0]{\spacefactor3000\relax}%
\providecommand \BibitemShut  [1]{\csname bibitem#1\endcsname}%
\let\auto@bib@innerbib\@empty
%</preamble>
\bibitem [{\citenamefont {Leggett}(1975)}]{RevModPhys.47.331}%
  \BibitemOpen
  \bibfield  {author} {\bibinfo {author} {\bibfnamefont {A.~J.}\ \bibnamefont
  {Leggett}},\ }\href {\doibase 10.1103/RevModPhys.47.331} {\bibfield
  {journal} {\bibinfo  {journal} {Rev. Mod. Phys.}\ }\textbf {\bibinfo {volume}
  {47}},\ \bibinfo {pages} {331} (\bibinfo {year} {1975})}\BibitemShut
  {NoStop}%
\bibitem [{\citenamefont {Tou}\ \emph {et~al.}(1996)\citenamefont {Tou},
  \citenamefont {Kitaoka}, \citenamefont {Asayama}, \citenamefont {Kimura},
  \citenamefont {\ifmmode~\bar{O}\else \={O}\fi{}nuki}, \citenamefont
  {Yamamoto},\ and\ \citenamefont {Maezawa}}]{PhysRevLett.77.1374}%
  \BibitemOpen
  \bibfield  {author} {\bibinfo {author} {\bibfnamefont {H.}~\bibnamefont
  {Tou}}, \bibinfo {author} {\bibfnamefont {Y.}~\bibnamefont {Kitaoka}},
  \bibinfo {author} {\bibfnamefont {K.}~\bibnamefont {Asayama}}, \bibinfo
  {author} {\bibfnamefont {N.}~\bibnamefont {Kimura}}, \bibinfo {author}
  {\bibfnamefont {Y.}~\bibnamefont {\ifmmode~\bar{O}\else \={O}\fi{}nuki}},
  \bibinfo {author} {\bibfnamefont {E.}~\bibnamefont {Yamamoto}}, \ and\
  \bibinfo {author} {\bibfnamefont {K.}~\bibnamefont {Maezawa}},\ }\href
  {\doibase 10.1103/PhysRevLett.77.1374} {\bibfield  {journal} {\bibinfo
  {journal} {Phys. Rev. Lett.}\ }\textbf {\bibinfo {volume} {77}},\ \bibinfo
  {pages} {1374} (\bibinfo {year} {1996})}\BibitemShut {NoStop}%
\bibitem [{\citenamefont {Ishida}\ \emph {et~al.}(1998)\citenamefont {Ishida},
  \citenamefont {Mukuda}, \citenamefont {Kitaoka}, \citenamefont {Asayama},
  \citenamefont {Mao}, \citenamefont {Mori},\ and\ \citenamefont
  {Maeno}}]{Nature.396.658}%
  \BibitemOpen
  \bibfield  {author} {\bibinfo {author} {\bibfnamefont {K.}~\bibnamefont
  {Ishida}}, \bibinfo {author} {\bibfnamefont {H.}~\bibnamefont {Mukuda}},
  \bibinfo {author} {\bibfnamefont {Y.}~\bibnamefont {Kitaoka}}, \bibinfo
  {author} {\bibfnamefont {K.}~\bibnamefont {Asayama}}, \bibinfo {author}
  {\bibfnamefont {Z.~Q.}\ \bibnamefont {Mao}}, \bibinfo {author} {\bibfnamefont
  {Y.}~\bibnamefont {Mori}}, \ and\ \bibinfo {author} {\bibfnamefont
  {Y.}~\bibnamefont {Maeno}},\ }\href@noop {} {\bibfield  {journal} {\bibinfo
  {journal} {Nature (London)}\ }\textbf {\bibinfo {volume} {396}},\ \bibinfo
  {pages} {658} (\bibinfo {year} {1998})}\BibitemShut {NoStop}%
\bibitem [{\citenamefont {Mackenzie}\ and\ \citenamefont
  {Maeno}(2003)}]{RevModPhys.75.657}%
  \BibitemOpen
  \bibfield  {author} {\bibinfo {author} {\bibfnamefont {A.~P.}\ \bibnamefont
  {Mackenzie}}\ and\ \bibinfo {author} {\bibfnamefont {Y.}~\bibnamefont
  {Maeno}},\ }\href {\doibase 10.1103/RevModPhys.75.657} {\bibfield  {journal}
  {\bibinfo  {journal} {Rev. Mod. Phys.}\ }\textbf {\bibinfo {volume} {75}},\
  \bibinfo {pages} {657} (\bibinfo {year} {2003})}\BibitemShut {NoStop}%
\bibitem [{\citenamefont {Maeno}\ \emph {et~al.}(2012)\citenamefont {Maeno},
  \citenamefont {Kittaka}, \citenamefont {Nomura}, \citenamefont {Yonezawa},\
  and\ \citenamefont {Ishida}}]{JPSJ.81.011009}%
  \BibitemOpen
  \bibfield  {author} {\bibinfo {author} {\bibfnamefont {Y.}~\bibnamefont
  {Maeno}}, \bibinfo {author} {\bibfnamefont {S.}~\bibnamefont {Kittaka}},
  \bibinfo {author} {\bibfnamefont {T.}~\bibnamefont {Nomura}}, \bibinfo
  {author} {\bibfnamefont {S.}~\bibnamefont {Yonezawa}}, \ and\ \bibinfo
  {author} {\bibfnamefont {K.}~\bibnamefont {Ishida}},\ }\href {\doibase
  10.1143/JPSJ.81.011009} {\bibfield  {journal} {\bibinfo  {journal} {J. Phys.
  Soc. Jpn.}\ }\textbf {\bibinfo {volume} {81}},\ \bibinfo {pages} {011009}
  (\bibinfo {year} {2012})}\BibitemShut {NoStop}%
\bibitem [{\citenamefont {Mackenzie}\ \emph {et~al.}(2017)\citenamefont
  {Mackenzie}, \citenamefont {Scaffidi}, \citenamefont {Hicks},\ and\
  \citenamefont {Maeno}}]{NPJ.2.40}%
  \BibitemOpen
  \bibfield  {author} {\bibinfo {author} {\bibfnamefont {A.~P.}\ \bibnamefont
  {Mackenzie}}, \bibinfo {author} {\bibfnamefont {T.}~\bibnamefont {Scaffidi}},
  \bibinfo {author} {\bibfnamefont {C.~W.}\ \bibnamefont {Hicks}}, \ and\
  \bibinfo {author} {\bibfnamefont {Y.}~\bibnamefont {Maeno}},\ }\href
  {\doibase 10.1038/s41535-017-0045-4} {\bibfield  {journal} {\bibinfo
  {journal} {npj Quantum Mater.}\ }\textbf {\bibinfo {volume} {2}},\ \bibinfo
  {pages} {40} (\bibinfo {year} {2017})}\BibitemShut {NoStop}%
\bibitem [{\citenamefont {Aoki}\ and\ \citenamefont
  {Flouquet}(2012)}]{JPSJ.81.011003}%
  \BibitemOpen
  \bibfield  {author} {\bibinfo {author} {\bibfnamefont {D.}~\bibnamefont
  {Aoki}}\ and\ \bibinfo {author} {\bibfnamefont {J.}~\bibnamefont
  {Flouquet}},\ }\href {\doibase 10.1143/JPSJ.81.011003} {\bibfield  {journal}
  {\bibinfo  {journal} {J. Phys. Soc. Jpn.}\ }\textbf {\bibinfo {volume}
  {81}},\ \bibinfo {pages} {011003} (\bibinfo {year} {2012})}\BibitemShut
  {NoStop}%
\bibitem [{\citenamefont {Aoki}\ and\ \citenamefont
  {Flouquet}(2014)}]{JPSJ.83.061011}%
  \BibitemOpen
  \bibfield  {author} {\bibinfo {author} {\bibfnamefont {D.}~\bibnamefont
  {Aoki}}\ and\ \bibinfo {author} {\bibfnamefont {J.}~\bibnamefont
  {Flouquet}},\ }\href {\doibase 10.7566/JPSJ.83.061011} {\bibfield  {journal}
  {\bibinfo  {journal} {J. Phys. Soc. Jpn.}\ }\textbf {\bibinfo {volume}
  {83}},\ \bibinfo {pages} {061011} (\bibinfo {year} {2014})}\BibitemShut
  {NoStop}%
\bibitem [{\citenamefont {Huxley}\ \emph {et~al.}(2007)\citenamefont {Huxley},
  \citenamefont {Yates}, \citenamefont {L{\'e}vy},\ and\ \citenamefont
  {Sheikin}}]{JPSJ.76.051011}%
  \BibitemOpen
  \bibfield  {author} {\bibinfo {author} {\bibfnamefont {A.~D.}\ \bibnamefont
  {Huxley}}, \bibinfo {author} {\bibfnamefont {S.~J.~C.}\ \bibnamefont
  {Yates}}, \bibinfo {author} {\bibfnamefont {F.}~\bibnamefont {L{\'e}vy}}, \
  and\ \bibinfo {author} {\bibfnamefont {I.}~\bibnamefont {Sheikin}},\ }\href
  {\doibase 10.1143/JPSJ.76.051011} {\bibfield  {journal} {\bibinfo  {journal}
  {J. Phys. Soc. Jpn.}\ }\textbf {\bibinfo {volume} {76}},\ \bibinfo {pages}
  {051011} (\bibinfo {year} {2007})}\BibitemShut {NoStop}%
\bibitem [{\citenamefont {Hattori}\ \emph {et~al.}(2014)\citenamefont
  {Hattori}, \citenamefont {Ihara}, \citenamefont {Karube}, \citenamefont
  {Sugimoto}, \citenamefont {Ishida}, \citenamefont {Deguchi}, \citenamefont
  {Sato},\ and\ \citenamefont {Yamamura}}]{JPSJ.83.061012}%
  \BibitemOpen
  \bibfield  {author} {\bibinfo {author} {\bibfnamefont {T.}~\bibnamefont
  {Hattori}}, \bibinfo {author} {\bibfnamefont {Y.}~\bibnamefont {Ihara}},
  \bibinfo {author} {\bibfnamefont {K.}~\bibnamefont {Karube}}, \bibinfo
  {author} {\bibfnamefont {D.}~\bibnamefont {Sugimoto}}, \bibinfo {author}
  {\bibfnamefont {K.}~\bibnamefont {Ishida}}, \bibinfo {author} {\bibfnamefont
  {K.}~\bibnamefont {Deguchi}}, \bibinfo {author} {\bibfnamefont {N.~K.}\
  \bibnamefont {Sato}}, \ and\ \bibinfo {author} {\bibfnamefont
  {T.}~\bibnamefont {Yamamura}},\ }\href {\doibase 10.7566/JPSJ.83.061012}
  {\bibfield  {journal} {\bibinfo  {journal} {J. Phys. Soc. Jpn.}\ }\textbf
  {\bibinfo {volume} {83}},\ \bibinfo {pages} {061012} (\bibinfo {year}
  {2014})}\BibitemShut {NoStop}%
\bibitem [{\citenamefont {Huxley}(2015)}]{PhysicaC.514.368}%
  \BibitemOpen
  \bibfield  {author} {\bibinfo {author} {\bibfnamefont {A.~D.}\ \bibnamefont
  {Huxley}},\ }\href {\doibase https://doi.org/10.1016/j.physc.2015.02.026}
  {\bibfield  {journal} {\bibinfo  {journal} {Phys. C (Amsterdam, Neth.)}\
  }\textbf {\bibinfo {volume} {514}},\ \bibinfo {pages} {368} (\bibinfo {year}
  {2015})}\BibitemShut {NoStop}%
\bibitem [{\citenamefont {Hattori}\ \emph {et~al.}(2012)\citenamefont
  {Hattori}, \citenamefont {Ihara}, \citenamefont {Nakai}, \citenamefont
  {Ishida}, \citenamefont {Tada}, \citenamefont {Fujimoto}, \citenamefont
  {Kawakami}, \citenamefont {Osaki}, \citenamefont {Deguchi}, \citenamefont
  {Sato},\ and\ \citenamefont {Satoh}}]{PhysRevLett.108.066403}%
  \BibitemOpen
  \bibfield  {author} {\bibinfo {author} {\bibfnamefont {T.}~\bibnamefont
  {Hattori}}, \bibinfo {author} {\bibfnamefont {Y.}~\bibnamefont {Ihara}},
  \bibinfo {author} {\bibfnamefont {Y.}~\bibnamefont {Nakai}}, \bibinfo
  {author} {\bibfnamefont {K.}~\bibnamefont {Ishida}}, \bibinfo {author}
  {\bibfnamefont {Y.}~\bibnamefont {Tada}}, \bibinfo {author} {\bibfnamefont
  {S.}~\bibnamefont {Fujimoto}}, \bibinfo {author} {\bibfnamefont
  {N.}~\bibnamefont {Kawakami}}, \bibinfo {author} {\bibfnamefont
  {E.}~\bibnamefont {Osaki}}, \bibinfo {author} {\bibfnamefont
  {K.}~\bibnamefont {Deguchi}}, \bibinfo {author} {\bibfnamefont {N.~K.}\
  \bibnamefont {Sato}}, \ and\ \bibinfo {author} {\bibfnamefont
  {I.}~\bibnamefont {Satoh}},\ }\href {\doibase 10.1103/PhysRevLett.108.066403}
  {\bibfield  {journal} {\bibinfo  {journal} {Phys. Rev. Lett.}\ }\textbf
  {\bibinfo {volume} {108}},\ \bibinfo {pages} {066403} (\bibinfo {year}
  {2012})}\BibitemShut {NoStop}%
\bibitem [{\citenamefont {Wu}\ \emph {et~al.}(2017)\citenamefont {Wu},
  \citenamefont {Bastien}, \citenamefont {Taupin}, \citenamefont {Paulsen},
  \citenamefont {Howald}, \citenamefont {Aoki},\ and\ \citenamefont
  {Brison}}]{NatComm.8.14480}%
  \BibitemOpen
  \bibfield  {author} {\bibinfo {author} {\bibfnamefont {B.}~\bibnamefont
  {Wu}}, \bibinfo {author} {\bibfnamefont {G.}~\bibnamefont {Bastien}},
  \bibinfo {author} {\bibfnamefont {M.}~\bibnamefont {Taupin}}, \bibinfo
  {author} {\bibfnamefont {C.}~\bibnamefont {Paulsen}}, \bibinfo {author}
  {\bibfnamefont {L.}~\bibnamefont {Howald}}, \bibinfo {author} {\bibfnamefont
  {D.}~\bibnamefont {Aoki}}, \ and\ \bibinfo {author} {\bibfnamefont {J.-P.}\
  \bibnamefont {Brison}},\ }\href {\doibase 10.1038/ncomms14480} {\bibfield
  {journal} {\bibinfo  {journal} {Nat. Commun.}\ }\textbf {\bibinfo {volume}
  {8}},\ \bibinfo {pages} {14480} (\bibinfo {year} {2017})}\BibitemShut
  {NoStop}%
\bibitem [{\citenamefont {Tokunaga}\ \emph {et~al.}(2015)\citenamefont
  {Tokunaga}, \citenamefont {Aoki}, \citenamefont {Mayaffre}, \citenamefont
  {Kr\"amer}, \citenamefont {Julien}, \citenamefont {Berthier}, \citenamefont
  {Horvati\ifmmode~\acute{c}\else \'{c}\fi{}}, \citenamefont {Sakai},
  \citenamefont {Kambe},\ and\ \citenamefont {Araki}}]{PhysRevLett.114.216401}%
  \BibitemOpen
  \bibfield  {author} {\bibinfo {author} {\bibfnamefont {Y.}~\bibnamefont
  {Tokunaga}}, \bibinfo {author} {\bibfnamefont {D.}~\bibnamefont {Aoki}},
  \bibinfo {author} {\bibfnamefont {H.}~\bibnamefont {Mayaffre}}, \bibinfo
  {author} {\bibfnamefont {S.}~\bibnamefont {Kr\"amer}}, \bibinfo {author}
  {\bibfnamefont {M.-H.}\ \bibnamefont {Julien}}, \bibinfo {author}
  {\bibfnamefont {C.}~\bibnamefont {Berthier}}, \bibinfo {author}
  {\bibfnamefont {M.}~\bibnamefont {Horvati\ifmmode~\acute{c}\else
  \'{c}\fi{}}}, \bibinfo {author} {\bibfnamefont {H.}~\bibnamefont {Sakai}},
  \bibinfo {author} {\bibfnamefont {S.}~\bibnamefont {Kambe}}, \ and\ \bibinfo
  {author} {\bibfnamefont {S.}~\bibnamefont {Araki}},\ }\href {\doibase
  10.1103/PhysRevLett.114.216401} {\bibfield  {journal} {\bibinfo  {journal}
  {Phys. Rev. Lett.}\ }\textbf {\bibinfo {volume} {114}},\ \bibinfo {pages}
  {216401} (\bibinfo {year} {2015})}\BibitemShut {NoStop}%
\bibitem [{\citenamefont {Braithwaite}\ \emph {et~al.}(2018)\citenamefont
  {Braithwaite}, \citenamefont {Aoki}, \citenamefont {Brison}, \citenamefont
  {Flouquet}, \citenamefont {Knebel}, \citenamefont {Nakamura},\ and\
  \citenamefont {Pourret}}]{PhysRevLett.120.037001}%
  \BibitemOpen
  \bibfield  {author} {\bibinfo {author} {\bibfnamefont {D.}~\bibnamefont
  {Braithwaite}}, \bibinfo {author} {\bibfnamefont {D.}~\bibnamefont {Aoki}},
  \bibinfo {author} {\bibfnamefont {J.-P.}\ \bibnamefont {Brison}}, \bibinfo
  {author} {\bibfnamefont {J.}~\bibnamefont {Flouquet}}, \bibinfo {author}
  {\bibfnamefont {G.}~\bibnamefont {Knebel}}, \bibinfo {author} {\bibfnamefont
  {A.}~\bibnamefont {Nakamura}}, \ and\ \bibinfo {author} {\bibfnamefont
  {A.}~\bibnamefont {Pourret}},\ }\href {\doibase
  10.1103/PhysRevLett.120.037001} {\bibfield  {journal} {\bibinfo  {journal}
  {Phys. Rev. Lett.}\ }\textbf {\bibinfo {volume} {120}},\ \bibinfo {pages}
  {037001} (\bibinfo {year} {2018})}\BibitemShut {NoStop}%
\bibitem [{\citenamefont {Hassinger}\ \emph {et~al.}(2008)\citenamefont
  {Hassinger}, \citenamefont {Aoki}, \citenamefont {Knebel},\ and\
  \citenamefont {Flouquet}}]{JPSJ.77.073703}%
  \BibitemOpen
  \bibfield  {author} {\bibinfo {author} {\bibfnamefont {E.}~\bibnamefont
  {Hassinger}}, \bibinfo {author} {\bibfnamefont {D.}~\bibnamefont {Aoki}},
  \bibinfo {author} {\bibfnamefont {G.}~\bibnamefont {Knebel}}, \ and\ \bibinfo
  {author} {\bibfnamefont {J.}~\bibnamefont {Flouquet}},\ }\href {\doibase
  10.1143/JPSJ.77.073703} {\bibfield  {journal} {\bibinfo  {journal} {J. Phys.
  Soc. Jpn.}\ }\textbf {\bibinfo {volume} {77}},\ \bibinfo {pages} {073703}
  (\bibinfo {year} {2008})}\BibitemShut {NoStop}%
\bibitem [{\citenamefont {Slooten}\ \emph {et~al.}(2009)\citenamefont
  {Slooten}, \citenamefont {Naka}, \citenamefont {Gasparini}, \citenamefont
  {Huang},\ and\ \citenamefont {de~Visser}}]{PhysRevLett.103.097003}%
  \BibitemOpen
  \bibfield  {author} {\bibinfo {author} {\bibfnamefont {E.}~\bibnamefont
  {Slooten}}, \bibinfo {author} {\bibfnamefont {T.}~\bibnamefont {Naka}},
  \bibinfo {author} {\bibfnamefont {A.}~\bibnamefont {Gasparini}}, \bibinfo
  {author} {\bibfnamefont {Y.~K.}\ \bibnamefont {Huang}}, \ and\ \bibinfo
  {author} {\bibfnamefont {A.}~\bibnamefont {de~Visser}},\ }\href {\doibase
  10.1103/PhysRevLett.103.097003} {\bibfield  {journal} {\bibinfo  {journal}
  {Phys. Rev. Lett.}\ }\textbf {\bibinfo {volume} {103}},\ \bibinfo {pages}
  {097003} (\bibinfo {year} {2009})}\BibitemShut {NoStop}%
\bibitem [{\citenamefont {Bastien}\ \emph {et~al.}(2016)\citenamefont
  {Bastien}, \citenamefont {Braithwaite}, \citenamefont {Aoki}, \citenamefont
  {Knebel},\ and\ \citenamefont {Flouquet}}]{PhysRevB.94.125110}%
  \BibitemOpen
  \bibfield  {author} {\bibinfo {author} {\bibfnamefont {G.}~\bibnamefont
  {Bastien}}, \bibinfo {author} {\bibfnamefont {D.}~\bibnamefont
  {Braithwaite}}, \bibinfo {author} {\bibfnamefont {D.}~\bibnamefont {Aoki}},
  \bibinfo {author} {\bibfnamefont {G.}~\bibnamefont {Knebel}}, \ and\ \bibinfo
  {author} {\bibfnamefont {J.}~\bibnamefont {Flouquet}},\ }\href {\doibase
  10.1103/PhysRevB.94.125110} {\bibfield  {journal} {\bibinfo  {journal} {Phys.
  Rev. B}\ }\textbf {\bibinfo {volume} {94}},\ \bibinfo {pages} {125110}
  (\bibinfo {year} {2016})}\BibitemShut {NoStop}%
\bibitem [{\citenamefont {Huy}\ \emph {et~al.}(2008)\citenamefont {Huy},
  \citenamefont {de~Nijs}, \citenamefont {Huang},\ and\ \citenamefont
  {de~Visser}}]{PhysRevLett.100.077002}%
  \BibitemOpen
  \bibfield  {author} {\bibinfo {author} {\bibfnamefont {N.~T.}\ \bibnamefont
  {Huy}}, \bibinfo {author} {\bibfnamefont {D.~E.}\ \bibnamefont {de~Nijs}},
  \bibinfo {author} {\bibfnamefont {Y.~K.}\ \bibnamefont {Huang}}, \ and\
  \bibinfo {author} {\bibfnamefont {A.}~\bibnamefont {de~Visser}},\ }\href
  {\doibase 10.1103/PhysRevLett.100.077002} {\bibfield  {journal} {\bibinfo
  {journal} {Phys. Rev. Lett.}\ }\textbf {\bibinfo {volume} {100}},\ \bibinfo
  {pages} {077002} (\bibinfo {year} {2008})}\BibitemShut {NoStop}%
\bibitem [{\citenamefont {Mineev}(2010)}]{PhysRevB.81.180504}%
  \BibitemOpen
  \bibfield  {author} {\bibinfo {author} {\bibfnamefont {V.~P.}\ \bibnamefont
  {Mineev}},\ }\href {\doibase 10.1103/PhysRevB.81.180504} {\bibfield
  {journal} {\bibinfo  {journal} {Phys. Rev. B}\ }\textbf {\bibinfo {volume}
  {81}},\ \bibinfo {pages} {180504} (\bibinfo {year} {2010})}\BibitemShut
  {NoStop}%
\bibitem [{\citenamefont {Tada}\ \emph {et~al.}(2016)\citenamefont {Tada},
  \citenamefont {Takayoshi},\ and\ \citenamefont
  {Fujimoto}}]{PhysRevB.93.174512}%
  \BibitemOpen
  \bibfield  {author} {\bibinfo {author} {\bibfnamefont {Y.}~\bibnamefont
  {Tada}}, \bibinfo {author} {\bibfnamefont {S.}~\bibnamefont {Takayoshi}}, \
  and\ \bibinfo {author} {\bibfnamefont {S.}~\bibnamefont {Fujimoto}},\ }\href
  {\doibase 10.1103/PhysRevB.93.174512} {\bibfield  {journal} {\bibinfo
  {journal} {Phys. Rev. B}\ }\textbf {\bibinfo {volume} {93}},\ \bibinfo
  {pages} {174512} (\bibinfo {year} {2016})}\BibitemShut {NoStop}%
\bibitem [{\citenamefont {Manago}\ \emph {et~al.}(2019)\citenamefont {Manago},
  \citenamefont {Kitagawa}, \citenamefont {Ishida}, \citenamefont {Deguchi},
  \citenamefont {Sato},\ and\ \citenamefont {Yamamura}}]{PhysRevB.99.020506}%
  \BibitemOpen
  \bibfield  {author} {\bibinfo {author} {\bibfnamefont {M.}~\bibnamefont
  {Manago}}, \bibinfo {author} {\bibfnamefont {S.}~\bibnamefont {Kitagawa}},
  \bibinfo {author} {\bibfnamefont {K.}~\bibnamefont {Ishida}}, \bibinfo
  {author} {\bibfnamefont {K.}~\bibnamefont {Deguchi}}, \bibinfo {author}
  {\bibfnamefont {N.~K.}\ \bibnamefont {Sato}}, \ and\ \bibinfo {author}
  {\bibfnamefont {T.}~\bibnamefont {Yamamura}},\ }\href {\doibase
  10.1103/PhysRevB.99.020506} {\bibfield  {journal} {\bibinfo  {journal} {Phys.
  Rev. B}\ }\textbf {\bibinfo {volume} {99}},\ \bibinfo {pages} {020506}
  (\bibinfo {year} {2019})}\BibitemShut {NoStop}%
\bibitem [{\citenamefont {Canepa}\ \emph {et~al.}(1996)\citenamefont {Canepa},
  \citenamefont {Manfrinetti}, \citenamefont {Pani},\ and\ \citenamefont
  {Palenzona}}]{JAlloysCompd.234.225}%
  \BibitemOpen
  \bibfield  {author} {\bibinfo {author} {\bibfnamefont {F.}~\bibnamefont
  {Canepa}}, \bibinfo {author} {\bibfnamefont {P.}~\bibnamefont {Manfrinetti}},
  \bibinfo {author} {\bibfnamefont {M.}~\bibnamefont {Pani}}, \ and\ \bibinfo
  {author} {\bibfnamefont {A.}~\bibnamefont {Palenzona}},\ }\href {\doibase
  10.1016/0925-8388(95)02037-3} {\bibfield  {journal} {\bibinfo  {journal} {J.
  Alloys Compd.}\ }\textbf {\bibinfo {volume} {234}},\ \bibinfo {pages} {225}
  (\bibinfo {year} {1996})}\BibitemShut {NoStop}%
\bibitem [{\citenamefont {Ihara}\ \emph {et~al.}(2010)\citenamefont {Ihara},
  \citenamefont {Hattori}, \citenamefont {Ishida}, \citenamefont {Nakai},
  \citenamefont {Osaki}, \citenamefont {Deguchi}, \citenamefont {Sato},\ and\
  \citenamefont {Satoh}}]{PhysRevLett.105.206403}%
  \BibitemOpen
  \bibfield  {author} {\bibinfo {author} {\bibfnamefont {Y.}~\bibnamefont
  {Ihara}}, \bibinfo {author} {\bibfnamefont {T.}~\bibnamefont {Hattori}},
  \bibinfo {author} {\bibfnamefont {K.}~\bibnamefont {Ishida}}, \bibinfo
  {author} {\bibfnamefont {Y.}~\bibnamefont {Nakai}}, \bibinfo {author}
  {\bibfnamefont {E.}~\bibnamefont {Osaki}}, \bibinfo {author} {\bibfnamefont
  {K.}~\bibnamefont {Deguchi}}, \bibinfo {author} {\bibfnamefont {N.~K.}\
  \bibnamefont {Sato}}, \ and\ \bibinfo {author} {\bibfnamefont
  {I.}~\bibnamefont {Satoh}},\ }\href {\doibase 10.1103/PhysRevLett.105.206403}
  {\bibfield  {journal} {\bibinfo  {journal} {Phys. Rev. Lett.}\ }\textbf
  {\bibinfo {volume} {105}},\ \bibinfo {pages} {206403} (\bibinfo {year}
  {2010})}\BibitemShut {NoStop}%
\bibitem [{\citenamefont {Ohta}\ \emph {et~al.}(2010)\citenamefont {Ohta},
  \citenamefont {Hattori}, \citenamefont {Ishida}, \citenamefont {Nakai},
  \citenamefont {Osaki}, \citenamefont {Deguchi}, \citenamefont {Sato},\ and\
  \citenamefont {Satoh}}]{JPSJ.79.023707}%
  \BibitemOpen
  \bibfield  {author} {\bibinfo {author} {\bibfnamefont {T.}~\bibnamefont
  {Ohta}}, \bibinfo {author} {\bibfnamefont {T.}~\bibnamefont {Hattori}},
  \bibinfo {author} {\bibfnamefont {K.}~\bibnamefont {Ishida}}, \bibinfo
  {author} {\bibfnamefont {Y.}~\bibnamefont {Nakai}}, \bibinfo {author}
  {\bibfnamefont {E.}~\bibnamefont {Osaki}}, \bibinfo {author} {\bibfnamefont
  {K.}~\bibnamefont {Deguchi}}, \bibinfo {author} {\bibfnamefont {N.~K.}\
  \bibnamefont {Sato}}, \ and\ \bibinfo {author} {\bibfnamefont
  {I.}~\bibnamefont {Satoh}},\ }\href {\doibase 10.1143/JPSJ.79.023707}
  {\bibfield  {journal} {\bibinfo  {journal} {J. Phys. Soc. Jpn.}\ }\textbf
  {\bibinfo {volume} {79}},\ \bibinfo {pages} {023707} (\bibinfo {year}
  {2010})}\BibitemShut {NoStop}%
\bibitem [{\citenamefont {Manago}\ \emph {et~al.}(2018)\citenamefont {Manago},
  \citenamefont {Ishida},\ and\ \citenamefont {Aoki}}]{PhysRevB.97.075130}%
  \BibitemOpen
  \bibfield  {author} {\bibinfo {author} {\bibfnamefont {M.}~\bibnamefont
  {Manago}}, \bibinfo {author} {\bibfnamefont {K.}~\bibnamefont {Ishida}}, \
  and\ \bibinfo {author} {\bibfnamefont {D.}~\bibnamefont {Aoki}},\ }\href
  {\doibase 10.1103/PhysRevB.97.075130} {\bibfield  {journal} {\bibinfo
  {journal} {Phys. Rev. B}\ }\textbf {\bibinfo {volume} {97}},\ \bibinfo
  {pages} {075130} (\bibinfo {year} {2018})}\BibitemShut {NoStop}%
\bibitem [{\citenamefont {Tou}\ \emph {et~al.}(2005)\citenamefont {Tou},
  \citenamefont {Ishida},\ and\ \citenamefont {Kitaoka}}]{JPSJ.74.1245}%
  \BibitemOpen
  \bibfield  {author} {\bibinfo {author} {\bibfnamefont {H.}~\bibnamefont
  {Tou}}, \bibinfo {author} {\bibfnamefont {K.}~\bibnamefont {Ishida}}, \ and\
  \bibinfo {author} {\bibfnamefont {Y.}~\bibnamefont {Kitaoka}},\ }\href
  {\doibase 10.1143/JPSJ.74.1245} {\bibfield  {journal} {\bibinfo  {journal}
  {J. Phys. Soc. Jpn.}\ }\textbf {\bibinfo {volume} {74}},\ \bibinfo {pages}
  {1245} (\bibinfo {year} {2005})}\BibitemShut {NoStop}%
\bibitem [{\citenamefont {Deguchi}\ \emph {et~al.}(2010)\citenamefont
  {Deguchi}, \citenamefont {Osaki}, \citenamefont {Ban}, \citenamefont
  {Tamura}, \citenamefont {Simura}, \citenamefont {Sakakibara}, \citenamefont
  {Satoh},\ and\ \citenamefont {Sato}}]{JPSJ.79.083708}%
  \BibitemOpen
  \bibfield  {author} {\bibinfo {author} {\bibfnamefont {K.}~\bibnamefont
  {Deguchi}}, \bibinfo {author} {\bibfnamefont {E.}~\bibnamefont {Osaki}},
  \bibinfo {author} {\bibfnamefont {S.}~\bibnamefont {Ban}}, \bibinfo {author}
  {\bibfnamefont {N.}~\bibnamefont {Tamura}}, \bibinfo {author} {\bibfnamefont
  {Y.}~\bibnamefont {Simura}}, \bibinfo {author} {\bibfnamefont
  {T.}~\bibnamefont {Sakakibara}}, \bibinfo {author} {\bibfnamefont
  {I.}~\bibnamefont {Satoh}}, \ and\ \bibinfo {author} {\bibfnamefont {N.~K.}\
  \bibnamefont {Sato}},\ }\href {\doibase 10.1143/JPSJ.79.083708} {\bibfield
  {journal} {\bibinfo  {journal} {J. Phys. Soc. Jpn.}\ }\textbf {\bibinfo
  {volume} {79}},\ \bibinfo {pages} {083708} (\bibinfo {year}
  {2010})}\BibitemShut {NoStop}%
\bibitem [{\citenamefont {Aoki}\ \emph {et~al.}(2009)\citenamefont {Aoki},
  \citenamefont {Matsuda}, \citenamefont {Taufour}, \citenamefont {Hassinger},
  \citenamefont {Knebel},\ and\ \citenamefont {Flouquet}}]{JPSJ.78.113709}%
  \BibitemOpen
  \bibfield  {author} {\bibinfo {author} {\bibfnamefont {D.}~\bibnamefont
  {Aoki}}, \bibinfo {author} {\bibfnamefont {T.~D.}\ \bibnamefont {Matsuda}},
  \bibinfo {author} {\bibfnamefont {V.}~\bibnamefont {Taufour}}, \bibinfo
  {author} {\bibfnamefont {E.}~\bibnamefont {Hassinger}}, \bibinfo {author}
  {\bibfnamefont {G.}~\bibnamefont {Knebel}}, \ and\ \bibinfo {author}
  {\bibfnamefont {J.}~\bibnamefont {Flouquet}},\ }\href {\doibase
  10.1143/JPSJ.78.113709} {\bibfield  {journal} {\bibinfo  {journal} {J. Phys.
  Soc. Jpn.}\ }\textbf {\bibinfo {volume} {78}},\ \bibinfo {pages} {113709}
  (\bibinfo {year} {2009})}\BibitemShut {NoStop}%
\bibitem [{\citenamefont {Miyake}(2014)}]{JPSJ.83.053701}%
  \BibitemOpen
  \bibfield  {author} {\bibinfo {author} {\bibfnamefont {K.}~\bibnamefont
  {Miyake}},\ }\href {\doibase 10.7566/JPSJ.83.053701} {\bibfield  {journal}
  {\bibinfo  {journal} {J. Phys. Soc. Jpn.}\ }\textbf {\bibinfo {volume}
  {83}},\ \bibinfo {pages} {053701} (\bibinfo {year} {2014})}\BibitemShut
  {NoStop}%
\bibitem [{\citenamefont {Fujimori}\ \emph {et~al.}(2015)\citenamefont
  {Fujimori}, \citenamefont {Ohkochi}, \citenamefont {Kawasaki}, \citenamefont
  {Yasui}, \citenamefont {Takeda}, \citenamefont {Okane}, \citenamefont
  {Saitoh}, \citenamefont {Fujimori}, \citenamefont {Yamagami}, \citenamefont
  {Haga}, \citenamefont {Yamamoto},\ and\ \citenamefont {\ifmmode~\bar{O}\else
  \={O}\fi{}nuki}}]{PhysRevB.91.174503}%
  \BibitemOpen
  \bibfield  {author} {\bibinfo {author} {\bibfnamefont {S.-i.}\ \bibnamefont
  {Fujimori}}, \bibinfo {author} {\bibfnamefont {T.}~\bibnamefont {Ohkochi}},
  \bibinfo {author} {\bibfnamefont {I.}~\bibnamefont {Kawasaki}}, \bibinfo
  {author} {\bibfnamefont {A.}~\bibnamefont {Yasui}}, \bibinfo {author}
  {\bibfnamefont {Y.}~\bibnamefont {Takeda}}, \bibinfo {author} {\bibfnamefont
  {T.}~\bibnamefont {Okane}}, \bibinfo {author} {\bibfnamefont
  {Y.}~\bibnamefont {Saitoh}}, \bibinfo {author} {\bibfnamefont
  {A.}~\bibnamefont {Fujimori}}, \bibinfo {author} {\bibfnamefont
  {H.}~\bibnamefont {Yamagami}}, \bibinfo {author} {\bibfnamefont
  {Y.}~\bibnamefont {Haga}}, \bibinfo {author} {\bibfnamefont {E.}~\bibnamefont
  {Yamamoto}}, \ and\ \bibinfo {author} {\bibfnamefont {Y.}~\bibnamefont
  {\ifmmode~\bar{O}\else \={O}\fi{}nuki}},\ }\href {\doibase
  10.1103/PhysRevB.91.174503} {\bibfield  {journal} {\bibinfo  {journal} {Phys.
  Rev. B}\ }\textbf {\bibinfo {volume} {91}},\ \bibinfo {pages} {174503}
  (\bibinfo {year} {2015})}\BibitemShut {NoStop}%
\bibitem [{\citenamefont {Brando}\ \emph {et~al.}(2016)\citenamefont {Brando},
  \citenamefont {Belitz}, \citenamefont {Grosche},\ and\ \citenamefont
  {Kirkpatrick}}]{RevModPhys.88.025006}%
  \BibitemOpen
  \bibfield  {author} {\bibinfo {author} {\bibfnamefont {M.}~\bibnamefont
  {Brando}}, \bibinfo {author} {\bibfnamefont {D.}~\bibnamefont {Belitz}},
  \bibinfo {author} {\bibfnamefont {F.~M.}\ \bibnamefont {Grosche}}, \ and\
  \bibinfo {author} {\bibfnamefont {T.~R.}\ \bibnamefont {Kirkpatrick}},\
  }\href {\doibase 10.1103/RevModPhys.88.025006} {\bibfield  {journal}
  {\bibinfo  {journal} {Rev. Mod. Phys.}\ }\textbf {\bibinfo {volume} {88}},\
  \bibinfo {pages} {025006} (\bibinfo {year} {2016})}\BibitemShut {NoStop}%
\bibitem [{\citenamefont {Taupin}\ \emph {et~al.}(2014)\citenamefont {Taupin},
  \citenamefont {Howald}, \citenamefont {Aoki},\ and\ \citenamefont
  {Brison}}]{PhysRevB.90.180501}%
  \BibitemOpen
  \bibfield  {author} {\bibinfo {author} {\bibfnamefont {M.}~\bibnamefont
  {Taupin}}, \bibinfo {author} {\bibfnamefont {L.}~\bibnamefont {Howald}},
  \bibinfo {author} {\bibfnamefont {D.}~\bibnamefont {Aoki}}, \ and\ \bibinfo
  {author} {\bibfnamefont {J.-P.}\ \bibnamefont {Brison}},\ }\href {\doibase
  10.1103/PhysRevB.90.180501} {\bibfield  {journal} {\bibinfo  {journal} {Phys.
  Rev. B}\ }\textbf {\bibinfo {volume} {90}},\ \bibinfo {pages} {180501(R)}
  (\bibinfo {year} {2014})}\BibitemShut {NoStop}%
\bibitem [{\citenamefont {Tou}\ \emph {et~al.}(1998)\citenamefont {Tou},
  \citenamefont {Kitaoka}, \citenamefont {Ishida}, \citenamefont {Asayama},
  \citenamefont {Kimura}, \citenamefont {{\=O}nuki}, \citenamefont {Yamamoto},
  \citenamefont {Haga},\ and\ \citenamefont {Maezawa}}]{PhysRevLett.80.3129}%
  \BibitemOpen
  \bibfield  {author} {\bibinfo {author} {\bibfnamefont {H.}~\bibnamefont
  {Tou}}, \bibinfo {author} {\bibfnamefont {Y.}~\bibnamefont {Kitaoka}},
  \bibinfo {author} {\bibfnamefont {K.}~\bibnamefont {Ishida}}, \bibinfo
  {author} {\bibfnamefont {K.}~\bibnamefont {Asayama}}, \bibinfo {author}
  {\bibfnamefont {N.}~\bibnamefont {Kimura}}, \bibinfo {author} {\bibfnamefont
  {Y.}~\bibnamefont {{\=O}nuki}}, \bibinfo {author} {\bibfnamefont
  {E.}~\bibnamefont {Yamamoto}}, \bibinfo {author} {\bibfnamefont
  {Y.}~\bibnamefont {Haga}}, \ and\ \bibinfo {author} {\bibfnamefont
  {K.}~\bibnamefont {Maezawa}},\ }\href {\doibase 10.1103/PhysRevLett.80.3129}
  {\bibfield  {journal} {\bibinfo  {journal} {Phys. Rev. Lett.}\ }\textbf
  {\bibinfo {volume} {80}},\ \bibinfo {pages} {3129} (\bibinfo {year}
  {1998})}\BibitemShut {NoStop}%
\end{thebibliography}%

\end{document}